\documentclass[pra,showpacs,twocolumn,superscriptaddress,amsmath,amssymb]{revtex4}

\usepackage{graphicx}
\usepackage{dcolumn}
\usepackage{bm}
\usepackage{epstopdf}
\usepackage{subfigure}
\usepackage{color}

\begin{document}


\title{Molecular ion trap-depletion spectroscopy of BaCl$^+$} 

\author{Kuang Chen}
\author{Steven J. Schowalter}
\affiliation{Department of Physics and Astronomy, University of California, Los Angeles, California 90095, USA}
\author{Svetlana Kotochigova}
\author{Alexander Petrov}
\altaffiliation{Alternative adress: St. Petersburg Nuclear Physics Institute, Gatchina, 188300; Department of Physics, St.Petersburg State University, 198904, Russia}
\affiliation{Department of Physics, Temple University, Philadelphia, Pennsylvania 19122, USA}
\author{Wade G. Rellergert}
\author{Scott T. Sullivan}
\author{Eric R. Hudson}
\email[Corresponding author: ]{eric.hudson@ucla.edu}
\affiliation{Department of Physics and Astronomy, University of California, Los Angeles, California 90095, USA}

\date{\today}

\begin{abstract}
We demonstrate a simple technique for molecular ion spectroscopy. BaCl$^+$
molecular ions are trapped in a linear Paul trap in the
presence of a room-temperature He buffer gas and photodissociated by
driving an electronic transition from the ground X$^1\Sigma^+$ state
to the repulsive wall of the A$^1\Pi$ state.  The photodissociation spectrum is recorded by monitoring
the induced trap loss of BaCl$^+$ ions as a function of excitation
wavelength. Accurate molecular potentials and spectroscopic constants are
determined. Comparison of the theoretical photodissociation cross-sections with the measurement shows excellent agreement. This study represents the first spectroscopic data for BaCl$^+$
and an important step towards the production of ultracold ground-state molecular ions.  \end{abstract}


\maketitle
Ultracold molecular ions in the rovibronic ground-state hold immense promise for fundamental research in physics and chemistry. Of particular interest are novel applications to quantum chemistry \cite{Willitsch2008}; a better understanding of interstellar cloud formation \cite{Smith1992} and the identification of potential carriers of the diffuse interstellar bands \cite{Reddy2010}; the implementation of scalable quantum computation architecture \cite{Schuster2009}; and precision measurement tests of fundamental physics \cite{Flambaum2007}. In pursuit of these goals, several groups have recently initiated work \cite{Hudson2009, Schneider2010, Staanum2010,Tong2010} to realize samples of cold, absolute ground-state molecular ions. In fact, Refs. \cite{Schneider2010,Staanum2010,Tong2010} have already reported the demonstration of species-specific cooling methods to produce molecular ions in the lowest few rotational states.

While these molecular ion cooling efforts, which include ultracold atom sympathetic cooling \cite{Hudson2009}, rovibrational optical pumping \cite{Schneider2010, Staanum2010}, and
state-selective ionization \cite{Tong2010}, are diverse in approach, they share the common need for detailed spectroscopic understanding of diatomic ions.
\begin{figure}[t]
\resizebox{0.9\columnwidth}{!}{
    \includegraphics{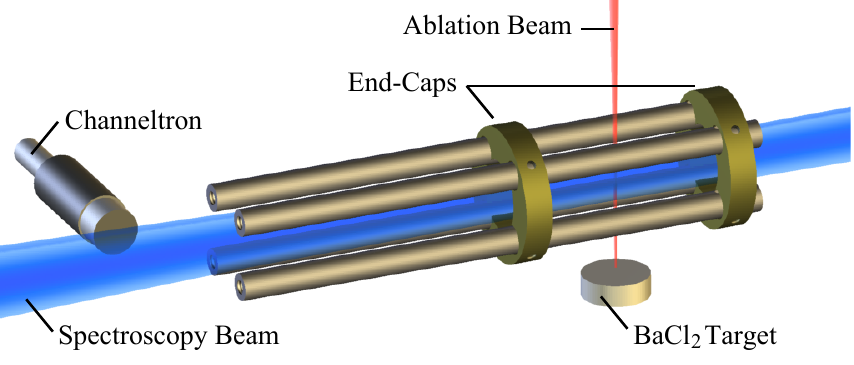}
}  \caption{(Color online) A schematic of the depletion spectroscopy apparatus based on a linear Paul trap.   \label{fig:app}}
\end{figure} 
However, compared to that of neutral molecules, spectroscopic data for molecular
ions is scarce. This can be attributed to the typically short lifetimes of
molecular ions due to fast ion-molecule reactions \cite{Saykally1981}. A
systematic review of the available spectroscopic data for simple diatomic
ions was carried out by Berkowitz and Groeneveld \cite{Berkowitz1983}
in 1983. In recent years, interest has shifted towards large molecular ions, atomic and molecular
clusters, and multiply charged ions \cite{Duncan2000}. Thus, for ultracold
molecular ion research to realize its full potential, a new effort in
small molecular ion spectroscopy is required.

Here, we report the use of a simple and general technique to record the first
spectroscopic data for BaCl$^+$ -- the molecular ion we have proposed
to cool in Ref.~\cite{Hudson2009}. As an ionically-bonded molecule
composed of two closed-shell atomic ions, Ba$^{2+}$ and Cl$^{-}$, BaCl$^+$
exhibits reduced chemical reactivity compared to other ions
and is
energetically forbidden from undergoing 2-body chemical reactions with the ultracold calcium
atoms used for sympathetic cooling \cite{Hudson2009}.
Further, its
large dipole moment and convenient rotational splitting are promising
for cavity QED experiments \cite{Schuster2009}. Thus, these results are important not only as the demonstration of a technique for recording molecular ion spectroscopy, but also as an important first step towards the use and application of a generic, robust method for the production of cold ground-state molecular ions.

Spectroscopic data is taken using a trap-depletion approach. BaCl$^+$ ions are trapped in a linear Paul trap in the presence of a room-temperature He
buffer gas and photodissociated by driving an electronic transition from
the ground X$^1\Sigma^+$ state to the repulsive wall of the A$^1\Pi$ state.
Ion trap parameters are carefully chosen to ensure that the photodissociation
products are not trapped. The photodissociation spectrum is then
recorded by simply monitoring the induced ion trap loss as a function
of excitation energy. This technique, which we estimate should be
easily applicable whenever the photofragments' mass-to-charge ratio differs by
$\geq$15\% from the parent molecular ion, may be a simple alternative
to both photofragment mass spectrometry \cite{Svendsen2010} and storage
ring based photodissociation spectroscopy \cite{Hechtfischer2007}.

In the remainder of this manuscript, we first give a detailed description of the experimental apparatus and technique. This is followed by the presentation of the first spectroscopic data for the BaCl$^+$ molecular ion. We continue with an {\it ab~initio} calculation of the BaCl$^+$ molecular structure and thermalized photodissociation spectra, which we compare to the experimental observation. 

Our apparatus, shown in Fig.~\ref{fig:app}, consists of a linear 
Paul trap housed in a vacuum chamber with a background gas pressure of $10^{-8}$~mbar. The ion trap is designed to allow radial loading of ions via laser ablation of a solid target, axial ejection of trapped ions into a channel electron multiplier for ion detection, and axial optical access for a spectroscopy beam. The ratio of the electrode radius $r_e$ to field radius $r_0$ is $r_e/r_0=0.401$. A pressed, annealed target of BaCl$_2$ mounted below the ion trap is ablated by a $\sim$1~mJ, 10~ns pulse of $1064$~nm laser radiation to create BaCl$^+$ molecular ions, which are trapped via the technique presented in Ref. \cite{Hashimoto2006}. A sample of Yb is mounted alongside the BaCl$_2$ target and is ablated to produce and trap Yb$^+$ ions, which are used as a control (described later).  A leak valve is used to insert up to $10^{-3}$~mbar of He buffer gas into the chamber to enhance the trapping of high-energy ablated ions through sympathetic cooling. The spectroscopy beam is generated by a frequency-doubled pulse dye laser (PDL) capable of photon energies up to $49,000$~cm$^{-1}$ with pulse energies of $\sim$1 mJ at a $10$~Hz  repetition rate.

Ablation is a complicated process that creates a plume of atoms, molecules, and clusters in various charge states \cite{Henyk2002}, all of which can potentially be loaded into the ion trap. For this experiment it is critical that BaCl$^+$ is the only ablation product stable in the trap. Further, we require that the possible photodissociation products, \textit{i.e.} Ba and Cl ions, are not trapped. To ensure the exclusivity of the trapping process, we record ion trap stability as a function of trap radiofrequency (rf) voltage and dc offset voltage at a rf driving frequency of $\Omega=2\pi\times200$~kHz. Based on the measurement of the trapped ion signal and comparison to theory \cite{Douglas2005}, trapping parameters are chosen outside the stability region of Ba$^+$. The presence of singly-charged ions heavier than BaCl$^+$ is ruled out by the lack of significant ion detection when operating the trap for heavier ions. To further confirm the exclusivity of BaCl$^+$ in the trap, the resonant excitation spectra \cite{Douglas2005} of BaCl$^+$ and Yb$^+$, both with mass to charge ratio of $\sim$173, are recorded and compared.
\begin{figure}[t]
\resizebox{0.9\columnwidth}{!}{
    \includegraphics{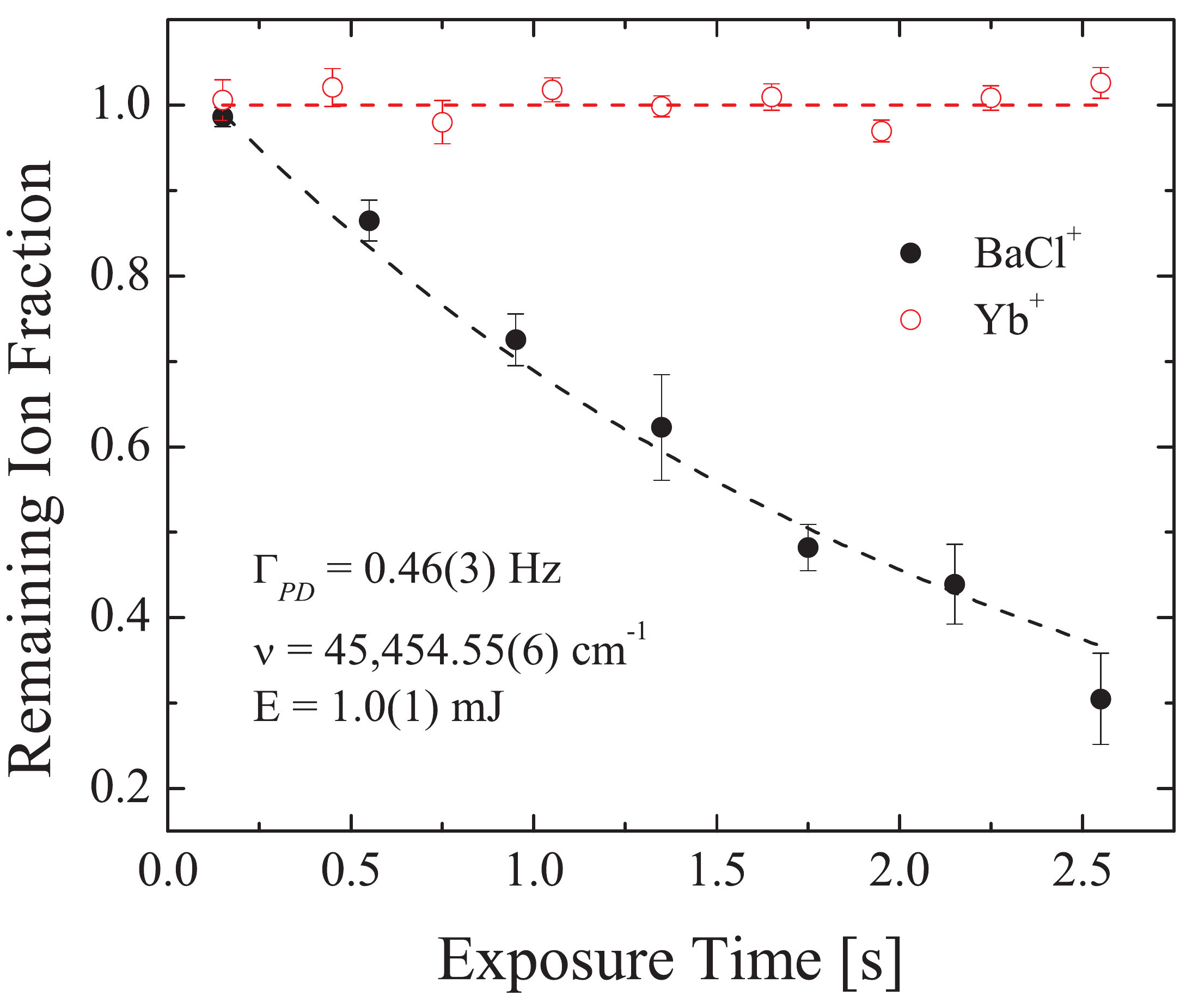}
}  \caption{(Color online) The effect of the PDL light on BaCl$^+$ and Yb$^+$. Each data point represents the mean of eight measurements with error bars reflecting the corresponding standard error. A single exponential decay curve is fit to the BaCl$^+$ data to calculate the photodissociation rate.\label{fig:decay}}
\end{figure}

This work contains the only spectroscopic information for BaCl$^+$ beyond a prior estimate of the ground-state dissociation energy, $D_e$, from a ligand field theory calculation \cite{Kaledin1998}, which suggests $D_e \approx -38,500$~cm$^{-1}$. Using this energy as the minimum direct dissociation energy, we search for photodissociation using trap-depletion spectroscopy. During each measurement, BaCl$^+$ ions are loaded into the trap via ablation, where collisions with the He buffer gas cool the ions' translational motion to $\sim$300~K in $\leq$~1~ms based on classical collision theory \cite{banks}.  Next, the spectroscopy beam is unshuttered potentially photodissociating the trapped BaCl$^+$ ions into untrapped Ba$^+$ and Cl atom fragments. After a variable exposure time, the spectroscopy beam is reshuttered and the remaining ions are axially ejected from the trap by grounding one of the end-cap electrodes. The ejected ions are then detected by a channel electron multiplier. The resulting ion signal is normalized to a control signal, for which the spectroscopy beam is always shuttered, in order to measure the fraction of ions remaining for a given exposure time. By repeating this procedure for a range of exposure times, a decay profile is measured and fit to a single exponential decay curve to obtain the photodissociation rate, $\Gamma_{PD}$. Typical data is shown in Fig.~\ref{fig:decay}.

To rule out systematic effects due to species-nonspecific loss processes, such as collisions with spectroscopy-beam-induced photoelectrons that result when scattered UV light impinges on the metallic trap electrodes, we also observe the spectroscopy beam's effect on Yb$^+$ as shown in Fig.~\ref{fig:decay}. For the range of photon energies used, we do not expect the PDL to induce a loss of Yb$^+$ through one-photon processes due to its high ionization potential, IP(Yb$^+$) = $98,207$~cm$^{-1}$~\cite{Sugar1979}. For all photon energies used, we do not observe any significant change in Yb$^+$ trap population, indicating the observed loss of BaCl$^+$ is species-specific.

\begin{figure}[t]
\resizebox{0.9\columnwidth}{!}{
    \includegraphics{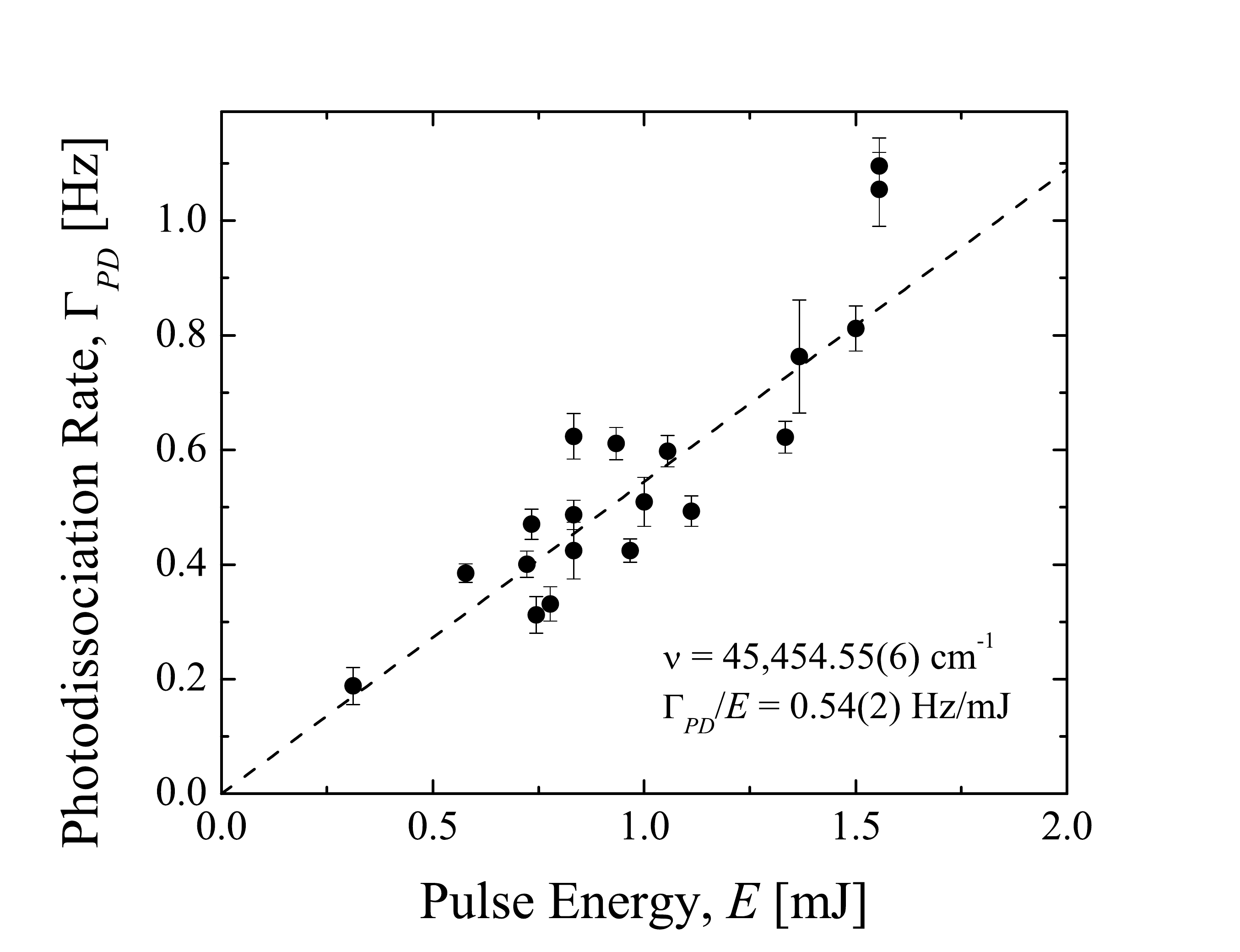}
}  \caption{Measured BaCl$^+$ photodissociation rate versus PDL pulse energy. The linear dependence is indicative of a one-photon photodissociation process. Each point is the result of a decay rate fit shown in Fig.~\ref{fig:decay} with error bars reflecting fitting error.    \label{fig:ratevenergy}}
\end{figure}

To rule out systematic effects due to species-specific loss processes other than direct photodissociation, such as multi-photon ionization of BaCl$^+$, we measure the dependence of $\Gamma_{PD}$ on the pulse energy and find a strong, linear relationship, shown in Fig.~\ref{fig:ratevenergy}, indicative of a one-photon process.  This data, combined with the results of our \textit{ab initio} calculations, confirms that the observed loss is due to single-photon photodissociation through the A$^1\Pi\leftarrow$X$^1\Sigma^+$ transition.

\begin{figure}[t]
\resizebox{0.9\columnwidth}{!}{
    \includegraphics{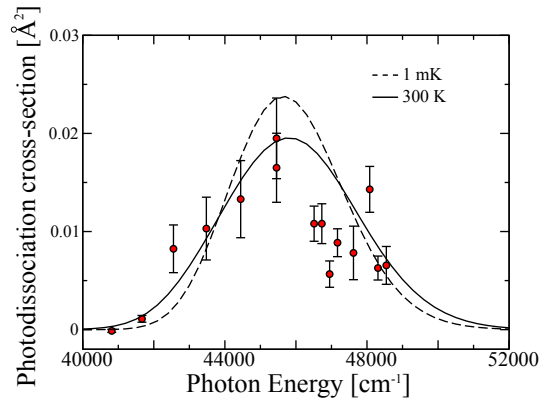}
}  \caption{(Color online) Experimental and theoretical cross-section values for the A$^1\Pi\leftarrow$X$^1\Sigma^+$ transition as functions of the photon energy. Each data point is associated with set of data similar to that in Fig.~\ref{fig:ratevenergy}.  The theoretical curves are thermally averaged for temperatures of 300 K and 1 mK.\label{fig:pdcs}}
\end{figure}

The photodissociation cross-section is calculated as
\begin{align}
\sigma_{PD}(\nu)=h \nu\frac{\Gamma_{PD}}{\overline{I}}=\frac{h\nu A}{r}\left(\frac{\Gamma_{PD}}{E}\right)\nonumber,
\end{align}
where $\Gamma_{PD}/E$ is the fitted slope in Fig.~\ref{fig:ratevenergy}, $A$ is the PDL elliptical beam area, and the average light intensity for the $r=10$~Hz repetition rate PDL is $\overline{I}= rE/A$. The measured photodissociation spectrum for a range of photon energies is shown in Fig.~\ref{fig:pdcs} alongside the results of our \textit{ab initio} calculation (described later).

Statistical error in the measurement of the photodissociation cross section arises from the measurement of the photodissociation rate and the average intensity of the PDL light. Error in the photodissociation rate is determined by a nonlinear fitting algorithm and is typically $<$~10$\%$. The error in the average intensity is experimentally manifested in the measurement of the PDL pulse energy ($\lesssim10\%$) and PDL beam area (typically $10-20\%$). Ultimately, the total statistical error for photodissociation cross-section values is  $<30\%$.

The majority of the systematic error in the measurement is attributed
to deviations from optimal overlap between the trapped ions and
the PDL light, which leads to systematically underestimating
the photodissociation rate.  To counter this problem, we ensure to
align the beam such that the photodissociation rate is maximized --
however, optimization is limited by the variation in ablation
yield. 
We estimate a total systematic error $\leq$40\%.

\begin{figure}[t]
 \includegraphics[scale=0.3]{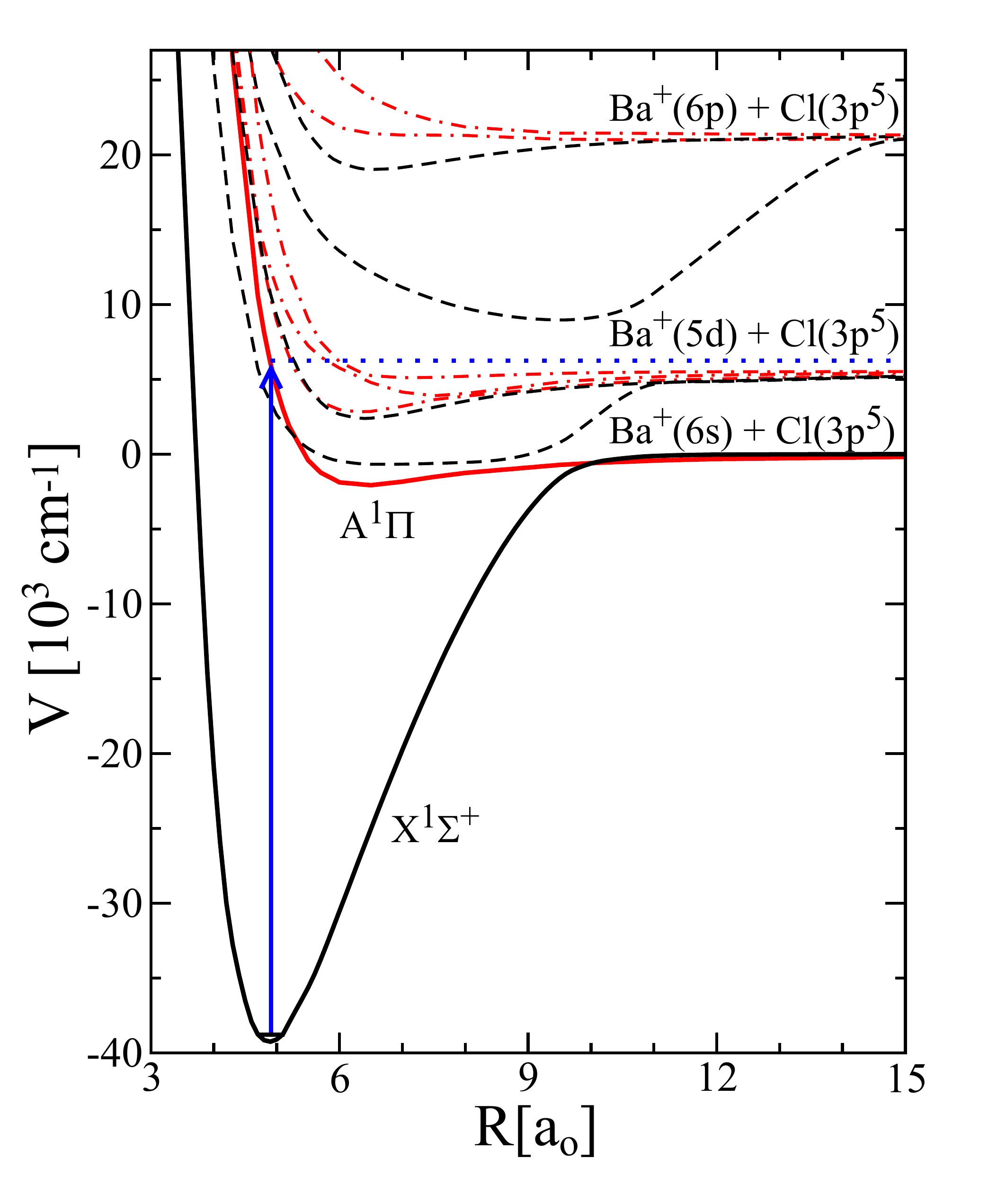}
 \caption{(Color online) Potential energy curves of the BaCl$^+$ molecular ion as a function of the internuclear separation $R$.  Solid, labeled curves indicate the potentials that are involved in the photodissociation scheme. Other $^1\Sigma^+$ ($^1\Pi$) potentials are shown by dashed (dash-dotted) lines. The vertical arrow indicates the A$^1\Pi\leftarrow$X$^1\Sigma^+$ single-photon photodissociation transition. \label{curves}}
\end{figure}

To interpret the recorded photodissociation spectrum we have calculated the ground and lowest excited electronic potentials of the BaCl$^+$ molecular ion, see Fig.~\ref{curves}, using a non-relativistic multi-configurational second-order perturbation theory (CASPT2) implemented in the MOLCAS software suite \cite{Karlstrom2003}.
The potential curves relevant to the experimental observation have solid lines and
are labeled X$^1\Sigma^+$ and A$^1\Pi$ for the ground and
first excited states, respectively --  spectroscopic constants are given in Tab.~\ref{constants}.
Other $^1\Sigma$ (dashed lines) and $^1\Pi$ (dash-dotted lines) potentials shown in Fig.~\ref{curves} are presented for completeness.  A detailed description of these and other symmetry potentials will be given in a forthcoming publication \cite{Kotochigova2011}.
\begin{table}[b]
\caption{BaCl$^+$ molecular spectroscopic constants} 
\begin{tabular}{lccccc}
\hline\\[-3mm]
State   \hspace{1mm}       & \hspace{3mm}  $R_e$  \hspace{3mm} & \hspace{3mm}$D_e$  \hspace{3mm} & \hspace{3mm} $\omega_e$ \hspace{3mm} & \hspace{2mm}      $\omega_e x_e$ \hspace{2mm} & $B_e$   \\
                     & [$a_0$] & [cm$^{-1}$] & [cm$^{-1}$] & [cm$^{-1}$]  & [cm$^{-1}$] \\[1mm]
\hline \\[-3mm] 
X$^1\Sigma^+$  & 4.85          & -39~055                &  328.3          & -1.56                     & 0.0918 \\ 
A$^1\Pi$              & 6.42           & -2~075                      &  90.53         &  -1.26                    & 0.0524\\ 
\hline
\end{tabular}
\label{constants}
\end{table}

Using these potentials we have developed a quantum mechanical model of the photodissociation process BaCl$^+$(X$^1\Sigma^+ , v J M)  + h\nu  \to$  BaCl$^+ $(A$^1\Pi, {\cal E}J'M') \to$ Ba$^+$(6s) + Cl(3p$^5$) with photon energy $h\nu$ and kinetic energy release $\cal E$. Based on the Franck-Condon principle, the absorption of a photon occurs for an internuclear separation where the kinetic energies in the initial and intermediate state are the same \cite{Gislason1972}. Fig.~\ref{curves} shows this separation by a vertical arrow for the $v=0$ vibrational level of the X$^1\Sigma^+$ state to the continuum or scattering
states of the A$^1\Pi$ potential, leading to a Ba$^+$ ion and Cl atom. For the photon energies used in this experiment, this is the only viable dissociation pathway. 
Thus, the photodissociation cross-section for each rovibrational level $vJ$ of the X$^1\Sigma^+$  state, assuming equal population in projections $M$ and linear photon polarization, is \cite{Williams1986}:
\begin{eqnarray}
\label{crosssection}
\sigma_{vJ}(\nu) &=&  4\pi^2 \alpha a_0^2 \,\frac{h\nu}{E_h}\, \frac{1}{2J+1 } \times \\
& &\sum_{J'MM'}  \frac{ |\langle A, {\cal E} J'M'  | d_z | X, v  J M \rangle |^2} {(e a_0/\sqrt{E_h})^2}\nonumber
\end{eqnarray}
where $E_h$ is the Hartree, $a_0$ is the Bohr radius, and $\alpha$ is the fine-structure constant.
The quantity $\langle A | d_z | X \rangle/(ea_0/\sqrt{E_h})$ is dimensionless and contains both
the radial and angular parts of the dipole moment.  To evaluate the dipole matrix element, we have used a MRCI electronic-structure method, developed in Ref.~\cite{KRb2003}.
For reference, at $R_e$ of the X$^1\Sigma^+$ state, the dipole moment is $\sim$0.2~$ea_0$, which is solely due to the Ba$^{2+}$Cl$^{-}$ character of the electronic wavefunction.

For comparison with experimental data, we thermally average the photodissociation cross-section as
\begin{equation}
    \sigma_{PD}(\nu) = \langle \sigma(\nu) \rangle_T =\frac{1}{Z} \sum_{vJ} (2J+1) \sigma_{vJ}(\nu) e^{-E_{vJ}/(kT)}\nonumber,
\end{equation}
where $Z$ and $k$ are the partition function and the Boltzmann constant, respectively. 
In Fig.~\ref{fig:pdcs}, the agreement between the experimentally observed  and the theoretical $T=300$~K photodissociation cross-section is shown to be good, indicating that the dissociation energy, dipole moment, and the slope of the A state potential at $R_e$ of the X state potential are accurate.

The thermalized photodissociation cross-section for a temperature of 1 mK is also presented in Fig.~\ref{fig:pdcs}, where a weak dependence of the cross-section on $T$ is evident. Thus, it would be difficult to use photodissociation spectroscopy to probe the internal molecular ion temperature.  However, with the experimentally verified molecular potentials we have identified a strong predissociation channel between the first excited $^1\Sigma$ and A$^1\Pi$ states, which we are now investigating. It is expected that the rovibrational resolution afforded by predissociation spectroscopy will allow us to efficiently measure molecular ion rovibrational temperatures \cite{frey1982}, a crucial step in the demonstration of the method proposed in Ref. \cite{Hudson2009}.

In conclusion, we have demonstrated a simple technique for molecular ion
trap-depletion spectroscopy and used it to record the first experimental spectroscopic
data for BaCl$^+$. We have also performed \textit{ab initio} calculations
of BaCl$^+$ structure and found good agreement with experimental
data. From these results, we have reported the first spectroscopic
constants for BaCl$^+$ and suggested assignments for the BaCl$^+$
molecular potentials. Finally, as BaCl$^+$ is a leading candidate
for ultracold molecular ion experiments, this work represents a necessary
step towards these important goals.  

This work was supported by NSF, ARO and
MURI-AFOSR on Polar Molecules Grants Nos. PHY-0855683, PHY-1005453 and
W911NF-10-1-0505.

\bibliography{BaCl+PDBib}
\end{document}